\documentclass[preprint,12pt]{elsarticle}

\pdfoutput=1
\usepackage{graphicx}
\usepackage{amssymb}
\usepackage{amsmath}
\usepackage{mwe} 
\usepackage{caption}
\usepackage{hyperref}

\usepackage[all]{hypcap}    
\usepackage[outdir=./]{epstopdf}

\journal{International Journal of Mass Spectrometry}

\begin{document}

\begin{frontmatter}


\title{Novel radio-frequency ion trap with spherical geometry}



\author{Houshyar Noshad}

\address{Department of Physics and Energy Engineering, Amirkabir University of Technology (Tehran Polytechnic), Tehran, Iran}

\author{Mostafa Honari-Latifpour}

\address{Department of Electrical Engineering, Amirkabir University of Technology (Tehran Polytechnic), Tehran, Iran}

\begin{abstract}
Confinement of single ions in a novel radio-frequency (RF) quadrupole ion trap with spherical shape is investigated. An optimization of this spherical ion trap (SIT) is carried out in order to suppress its nonlinearity substantially by eliminating the electric octupole moment. Hence, a trapping potential and consequently an electric field very similar to the ideal quadrupole ion trap (QIT) are obtained. Afterwards, three stability regions for the optimized SIT are numerically computed. The regions coincide well with those reported in the literature for the ideal QIT. The reason is attributed to the zero electric octupole moment of our proposed trap. The SIT’s simple geometry and relative ease of fabrication along with its increased trapping volume compared to the conventional hyperbolic quadrupole ion trap, make it an appropriate choice for miniaturization.
\end{abstract}

\begin{keyword}
Ion Trap \sep Nonlinearity \sep Multipole Fields


\end{keyword}

\end{frontmatter}


\section{Introduction}
\label{S:1}

Radio-frequency ion traps are frequently used for purposes ranging from mass spectrometry \cite{march_introduction_1997} to quantum information processing \cite{cirac_quantum_1995} \cite{haffner_quantum_2008} \cite{schmidt_spectroscopy_2005}, ultra-precise atomic clock sand atomic physics \cite{rosenband_frequency_2008}\cite{douglas_linear_2005}. In the context of mass spectrometry, RF traps are utilized to separate particles based on their charge to mass ratio \cite{louris_instrumentation_1987}\cite{seidelin_microfabricated_2006}. Quadrupole ion trap with hyperbolic electrodes, the so-called Paul traps, are a promising candidate for the realization of scalable quantum information processing devices \cite{kielpinski_architecture_2002}. For the implementation to be able to execute the large-scale quantum algorithms, this scheme requires trapping and manipulation of many ions as quantum bits, and thus, a small trap size \cite{wesenberg_electrostatics_2008}. Additionally, in the community of mass spectrometry for reasons such as those discussed in Ref. \cite{ouyang_rectilinear_2004}, there is a growing interest in miniature mass spectrometers. In order to miniaturize the trap, it is very important for the device to have a simple geometry to make the fabrication feasible. This is satisfied for our proposed ion trap in contrast to a conventional quadrupole Paul trap.

As another disadvantage of a practical quadrupole QIT, one can see the appearance of the higher order electric multipole components \cite{gupta_numerical_2008} inside the trap, which is attributed to the truncation of the hyperbolic-shaped electrodes. These higher order fields are known to have a surprisingly strong effect on the operation of the traps \cite{sudakov_effective_2001}. In our proposed spherical ion trap, the electric octupole component inside the sphere has been eliminated by the optimization of the trap. Hence, the electric quadrupole component is the dominant term of the electric field inside the trap.

It is worthwhile to note that, the scientists were first interested in a trap in which the confined ions could oscillate as the harmonic oscillator, namely similar to oscillation of an ideal spring. This motivation led to a pure electric quadrupole field inside the trap, and consequently hyperbolic shapes for the electrodes of the trap. In other words, scientist’s interest in a simple oscillation of trapped ions led to a type of trap with hyperbolic-shaped electrodes, which was named as Paul trap. Hence, due to the simple form of the ion oscillation, the equations of motion for the ions in the trap were obtained as the well-known Mathieu differential equation.

In this article, we present a novel spherical ion trap with azimuthal symmetric electrodes. The electric potential is analytically obtained, and the equations of motion for the charged particles inside the trap are demonstrated to be of the standard Mathieu form. Classical equations of motion for a particle in the trap are numerically solved, and three stability regions in the plane defined by Mathieu parameters are computed. By optimizing the geometry of the trap we achieve a potential field very similar to pure quadrupolar form. This trap has a simpler fabrication compared to a hyperbolic Paul trap, and consequently it is more amenable to miniaturization.




\begin{figure}[h]
\centering\includegraphics[width=1\linewidth]{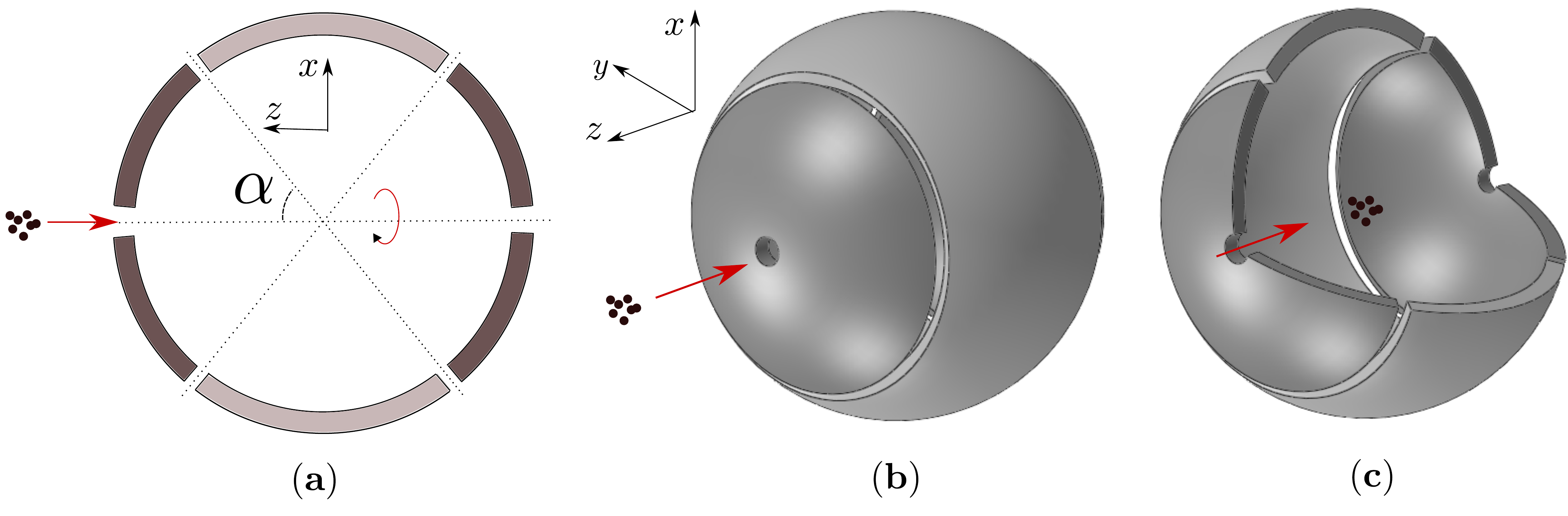}
\caption{The geometry of the spherical ion trap. The origin of the depicted coordinate systemis considered to coincide with the center of the trap. (a) The cross section of the proposed SIT in the x-z plane. As it is shown, the trap results from revolving this cross section around the z axis by 360 degrees. The two left and right electrodes (in dark gray) are driven by an RF potential and the middle electrode (in lighter gray) is grounded. (b) The full structure of the proposed trap. In addition to the invariance of the geometry under rotation around the z axis (rotational or azimuthal symmetry) the geometry is symmetric with respect to the x-y plane (left-right symmetry). (c) The confining point is the center of the sphere.}
\label{fig:1}
\end{figure}
\section{Trap Design}
\label{S:2}
The general geometry of the spherical ion trap is illustrated in Fig. \ref{fig:1}. All of the three electrodes are parts of a single sphere as is shown in the figure. The left and right caps are supplied with an RF potential and the middle electrode is grounded. The trap is left-right symmetric with respect to the x-y plane and also rotationally around the z axis (azimuthal symmetry). A consequence of the symmetric geometry of the electrodes is that the RF field exhibits a trapping point rather than a trapping line as in the linear Paul trap, and thus the confining fields originate merely from the RF potential and a trapping point at the center of the trap can be achieved without applying any DC voltage to the electrodes \cite{kim_surface-electrode_2010}. Furthermore, the symmetry of the trap with respect to the x-y plane, along with the azimuthal symmetry of the trap, causes the odd multipole terms of the potential to vanish and thus reduces the nonlinearities in the ion’s equations of motion that are caused by multipole components other than the quadrupole term. In addition to the analytical expression of the electric potential, we have solved the Laplace’s equation numerically in order to observe the behavior of the potential while taking the gaps between the electrodes into consideration. Existence of a potential saddle point at the center of the trap is shown through finite element analysis of the Laplace’s equation. Contour plots of the potential inside the trap are represented in Fig. \ref{fig:2}, which depicts a saddle point at the center of the trap. In fact, the existence of this saddle point causes RF traps to be able to confine charged particles dynamically \cite{major_charged_2006}. This conclusion is supported by the Earnshaw’s theorem as well \cite{griffiths_introduction_1999}.

This section includes three subsections as follows: \ref{sub:1} Derivation of an analytical expression for the electric potential inside the SIT. \ref{sub:2} Derivation of the equations of motion for an ion into the trap. \ref{sub:3} Optimization of the trap geometry to suppress the octupole term of the potential.

\subsection{Derivation of the Electric Potential}
\label{sub:1}
In order to obtain an analytical solution for the potential, we have modeled the trap as a spherical surface with radius $R$ and the following static surface potential:

\begin{equation}
  \Phi(R,\theta) =\begin{cases}
  0 \quad \quad \alpha<\theta<\pi-\alpha \\
  V_0 \quad \quad \quad  \text{otherwise}
  \end{cases}
  \label{eq:1}
\end{equation}

The potential $V_0$ has a form of $V_0=V\cos \Omega t$ , or more generally $V_0=U+V\cos \Omega t$ with $U$  being a DC term. The angle $\theta$  is measured starting from the z-axis as usual in spherical coordinates. The above potential means that the presence of the gaps were not considered in this analytical solution, although the gaps are accounted for in the FEM calculation of the potential (Fig. \ref{fig:2}). It should be noted that while in practical setups, an arrangement with RF-driven middle electrode and grounded caps is preferred for experimental reasons, only the potential difference between the caps and the middle electrodes is physically important for formation of a confining potential. As is indicated in Eq. \ref{eq:1}, in the calculations we considered the caps at $V_0$   instead of the central electrode when solving the problem analytically. Mathematically, the current boundary potential is equivalent to applying $-V_0$  to the central electrode and grounding the caps, which means a phase shift of $\pi$   compared to the mentioned experimental set up.

We will obtain the electric potential inside the trap expanded in terms of $\dfrac{r}{R}$ , where $r$ and $R$ are the radial distance and trap radius, respectively:

\begin{equation}
\Phi(r,\theta)=\sum_{n=0}^{\infty} A_n \left (\frac{r}{R}\right )^n P_n(\cos \theta)
\label{eq:2}
\end{equation}

Considering the potential in this form, known as multipole expansion, is of particular importance for our purpose. The reason is that it allows us to easily find our desired multipole components of the electric potential such as the quadrupole and octupole fields, that each have important effects on the performance of the trap. The $A_n$  coefficients are called multipole moments. We start by the well-known solution of the Laplace equation in spherical coordinates for azimuthal symmetric geometries \cite{jackson_classical_1962}

\begin{equation}
\Phi(r,\theta)=\sum_{n=0}^{\infty} B_nr^n P_n(\cos \theta)
\label{eq:3}
\end{equation}

In order to obtain a series solution for $B_n$ we have used the following expansion of Legendre polynomials

\begin{equation}
P_n(u)=\frac{1}{2^n}\sum_{m=0}^{[n/2]}\frac{(-1)^m(2n-2m)!}{m!(n-m)!(n-2m)!}u^{n-2m}
\label{eq:4}
\end{equation}

After some algebraic calculations, the coefficients are obtained by the following relation

\begin{equation}
  B_n=\begin{cases}
  V_0\frac{2n+1}{R^n}\sum_{m=0}^{n/2}(-1)^m\frac{(2n-2m)!(1-\cos^{n-2m+1}\alpha)}{2^nm!(n-m!)(n-2m+1)!}\quad \text{even } n \\
  0 \quad \quad \quad \quad \text{odd } n
  \end{cases}
  \label{eq:5}
\end{equation}

Thus, the potential becomes

\begin{equation}
  \begin{aligned}
  \Phi (r,\theta )=\sum\limits_{k=0}^{\infty }{{{B}_{2k}}}{{r}^{2k}}{{P}_{2k}}(\cos \theta )= \quad \quad \quad\quad \quad \quad \quad\quad\\
  \sum\limits_{k=0}^{\infty }{{{V}_{0}}}\frac{4k+1}{{{R}^{2k}}}\sum\limits_{m=0}^{k}{{{(-1)}^{m}}}\frac{(4k-2m)!(1-{{\cos }^{2k-2m+1}}\alpha )}{{{2}^{2k}}m!(2k-m)!(2k-2m+1)!}{{r}^{2k}}{{P}_{2k}}(\cos \theta )  
  \end{aligned}
  \label{eq:6}
\end{equation}

The following expression in the above series for the potential expression can be proved to be zero for all values of $k$:

\begin{figure}[h]
\centering\includegraphics[width=0.6\linewidth]{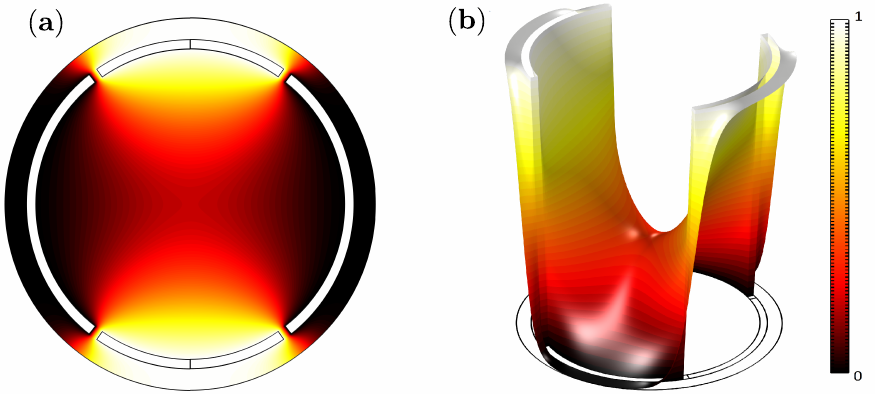}
\caption{(a) Contour plot of the potential viewed in a cross section of the SIT (x-z plane) (b) The same plot in (a) with the potential levels expressed as height to show the saddle point more clear}
\label{fig:2}
\end{figure}

\begin{equation}
\sum_{m=0}^k\frac{(-1)^m(4k-2m)!}{m!(2k-m)!(2k-2m+1)!}
\label{eq:7}
\end{equation}

Hence, the potential becomes

\begin{equation}
\Phi (r,\theta )=\sum_{k=0}^{\infty }{{{V}_{0}}}\left (\frac{4k+1}{{{R}^{2k}}}\sum\limits_{m=0}^{k}{{{(-1)}^{m+1}}}\frac{(4k-2m)!(1-{{\cos }^{2k-2m+1}}\alpha )}{{{2}^{2k}}m!(2k-m)!(2k-2m+1)!}\right ){{r}^{2k}}{{P}_{2k}}(\cos \theta )  
\label{eq:8}
\end{equation}

The above expansion can be written in the form of Eq. \ref{eq:1} with the multipole moments obtained as

\begin{equation}
A_{2k}=\frac{4k+1}{{{2}^{2k}}}\sum\limits_{m=0}^{k}{{{(-1)}^{m+1}}}\frac{(4k-2m)!(1-{{\cos }^{2k-2m+1}}\alpha )}{m!(2k-m)!(2k-2m+1)!}
\label{eq:9}
\end{equation}

As previously mentioned, the left-right symmetry of the trap geometry with respect to the x-y plane causes the odd terms of the multipole expansion to be zero.

\subsection{Derivation of the Equations of Motion}
\label{sub:2}

Due to the azimuthal symmetry of the trap, we are dealing with an inherently two dimensional problem and the ions’ trajectories inside the trap remain in one plane. Hence, for simplicity, we consider the ion motion to be in thex-z plane as shown in Fig. \ref{fig:1}, (a). So the potential becomes

\begin{equation}
  \begin{aligned}
  \Phi (x,z)=\sum\limits_{k=0}^{\infty }{{{A}_{2k}}}\frac{{{({{x}^{2}}+{{z}^{2}})}^{k}}}{{{R}^{2k}}}{{P}_{2k}}(\frac{z}{\sqrt{{{x}^{2}}+{{z}^{2}}}})= \quad \quad \\
  {{A}_{0}}+\frac{{{A}_{2}}}{{{R}^{2}}}(\frac{2{{z}^{2}}-{{x}^{2}}}{2})+\frac{{{A}_{4}}}{{{R}^{4}}}(\frac{3{{x}^{4}}-24{{x}^{2}}{{z}^{2}}+{{x}^{4}}}{8})+... 
  \end{aligned}
  \label{eq:10}
\end{equation}

The first non-constant term in the potential is the quadrupole component of the potential. This term leads to the standard Mathieu equations for the motion. To write the equations of motion in a concise form, we denote all higher order components of the potential collectively as $\dfrac{\Phi'(x,z)}{R^2}$

\begin{equation}
\Phi (x,z)={{A}_{0}}+{{A}_{2}}(\frac{2{{z}^{2}}-{{x}^{2}}}{2{{R}^{2}}})+\frac{\Phi '(x,z)}{{{R}^{2}}}
\label{eq:11}
\end{equation}

This way, we can maintain the Mathieu form of the equations of motion for this high-order non-quadratic potential. The higher order terms appear as perturbative terms to the standard Mathieu equation. Here we consider the potential applied to the central electrode in its most general form as $V_0=U+V\cos \Omega t$. Taking the gradient of the potential to obtain the electric field and drawing on the Newton’s second law of motion, classical equations of motion for a particle of mass $M$ and electric charge $Q$ inside the trap are obtained

\begin{equation}
\frac{{{d}^{2}}z}{d{{\tau }^{2}}}+({{a}_{z}}-2{{q}_{z}}\cos 2\tau )z=\frac{-1}{2{{A}_{2}}}({{a}_{z}}-2{{q}_{z}}\cos 2\tau )\frac{\partial \Phi '}{\partial z}
\label{eq:12}
\end{equation}

\begin{equation}
\frac{{{d}^{2}}x}{d{{\tau }^{2}}}+({{a}_{x}}-2{{q}_{x}}\cos 2\tau )z=\frac{-1}{2{{A}_{2}}}({{a}_{z}}-2{{q}_{z}}\cos 2\tau )\frac{\partial \Phi '}{\partial x}
\label{eq:13}
\end{equation}

In the relations, the dimensionless parameters  $a_x$, $a_z$, $q_x$ ,$q_z$ and $\tau$ are given by

\begin{equation}
\tau=\frac{\Omega t}{2}
\label{eq:14}
\end{equation}

\begin{equation}
a_z=-2a_x=\frac{8QUA_2}{MR^2\Omega ^2}
\label{eq:15}
\end{equation}

\begin{equation}
q_z=-2q_x=\frac{-4QVA_2}{MR^2\Omega ^2}
\label{eq:16}
\end{equation}

\begin{figure}[h]
\centering\includegraphics[width=.6\linewidth]{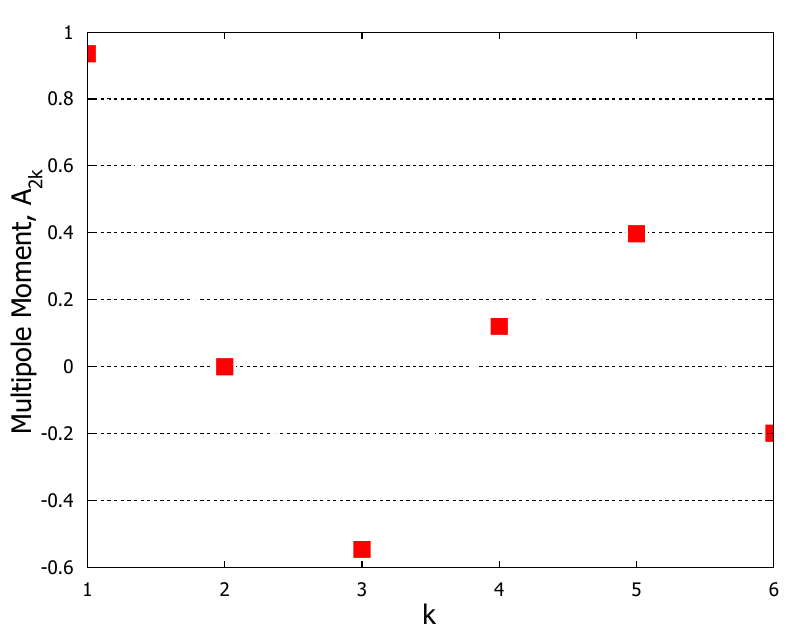}
\caption{Electric multipole moments of the potential expansion for a SIT with $\alpha \simeq {{49}^{\circ }}$ . Only these multipole components (${{A}_{2}},{{A}_{4}},{{A}_{6}},{{A}_{8}},{{A}_{10}},{{A}_{12}}$) of the potential were considered in numerical simulations and higher order terms were set to zero.}
\label{fig:3}
\end{figure}

It can be seen that in Eqs. \ref{eq:11} and \ref{eq:12}, the non-zero higher order multipole terms in the potential appear as perturbative terms to the standard Mathieu equation.

\subsection{Optimization of the Trap Geometry}
\label{sub:3}

Simplification of the electrode shapes introduces higher order multipole terms to the quadrupolar potential.The deviation of the potential from pure quadrupolar form leads to some noticeable effects. For example, nonlinear resonances can occur which can lead to possibly unwanted ejection of the ion from the trap \cite{wang_non-linear_1993}. The practically important higher order multipole terms are the hexapole ($A_3$), the octupole ($A_4$) and the dodecapole ($A_6$) terms \cite{plass_theory_2003}. The hexapole term is automatically zero due to the symmetry of the trap with respect to x-y plane. We can eliminate one of the two remaining terms by adjusting the angle   of the caps. Since the octupole term affects the potential more than the dodecapole, we choose this term for elimination. For this purpose, we equate the octupole moment $A_4$  from Eq. \ref{eq:8} to zero

\begin{equation}
{{A}_{4}}=\frac{9}{{{4}^{2}}}\sum\limits_{m=0}^{2}{{{(-1)}^{m+1}}}\frac{(8-2m)!({{\cos }^{4-2m+1}}\alpha )}{m!(4-m)!(4-2m+1)!}=0
\label{eq:17}
\end{equation}

\begin{figure}[h]
\centering\includegraphics[width=.8\linewidth]{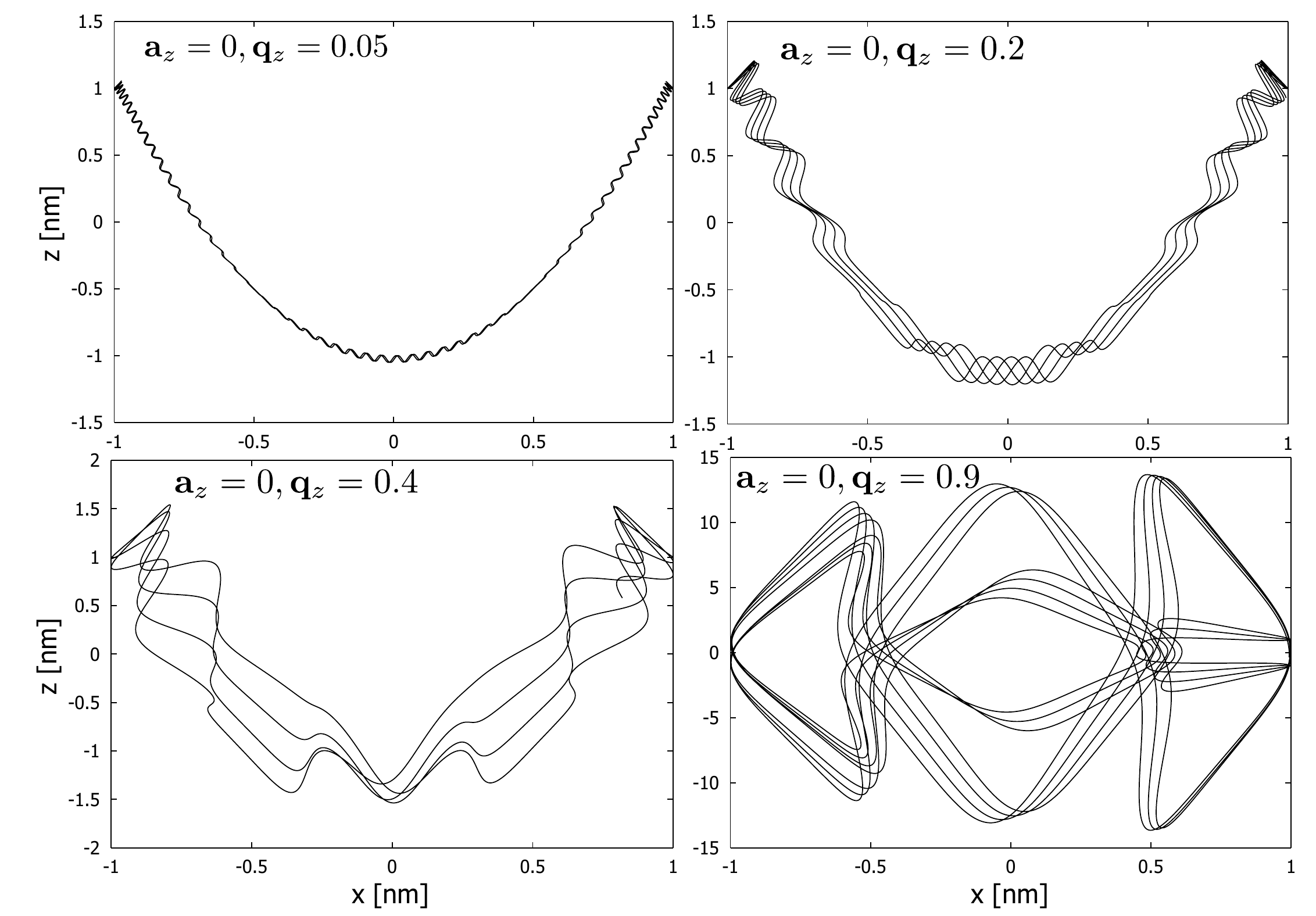}
\caption{Trajectories of the ion inside the trap for different Mathieu parameters. For small values of the micromotion is negligible (top, left). As the   parameter gets larger, the micromotion becomes comparable to the secular motion of the ion, and near the edge of the stability region (bottom, right) the micromotion amplitude is in the order of the secular motion \cite{sudakov_effective_2001}.}
\label{fig:4}
\end{figure}

The above equation yields

\begin{equation}
7{{\cos }^{5}}\alpha -10{{\cos }^{3}}\alpha +3\cos \alpha =0
\label{eq:18}
\end{equation}

Hence, the value of $\alpha$ which optimizes the spherical ion trap to suppress the electric octupole term is obtained by solving the above equation

\begin{equation}
\cos \alpha =\sqrt{\frac{3}{7}}\,\,\to \,\,\alpha \simeq {{49}^{\circ }}
\label{eq:19}
\end{equation}

With the octupole moment being zero, the next non-zero multipole term after the quadrupole term will be the dodecapole term $A_6$  with a $\dfrac{r^5}{R^6}$  dependence. This means that the region in which the field is approximately of pure quadrupole form expands considerably inside the optimized spherical ion trap. It is worth noting that the value of $\alpha \simeq {{49}^{\circ }}$  is selected in our computation in order to show excellent agreement between our results and those obtained for the Paul trap. The multipole moments corresponding to  $\alpha \simeq {{49}^{\circ }}$, which are used for the presented numerical calculations are shown in Fig. \ref{fig:3}.

\begin{figure}[h]
\centering\includegraphics[width=.75\linewidth]{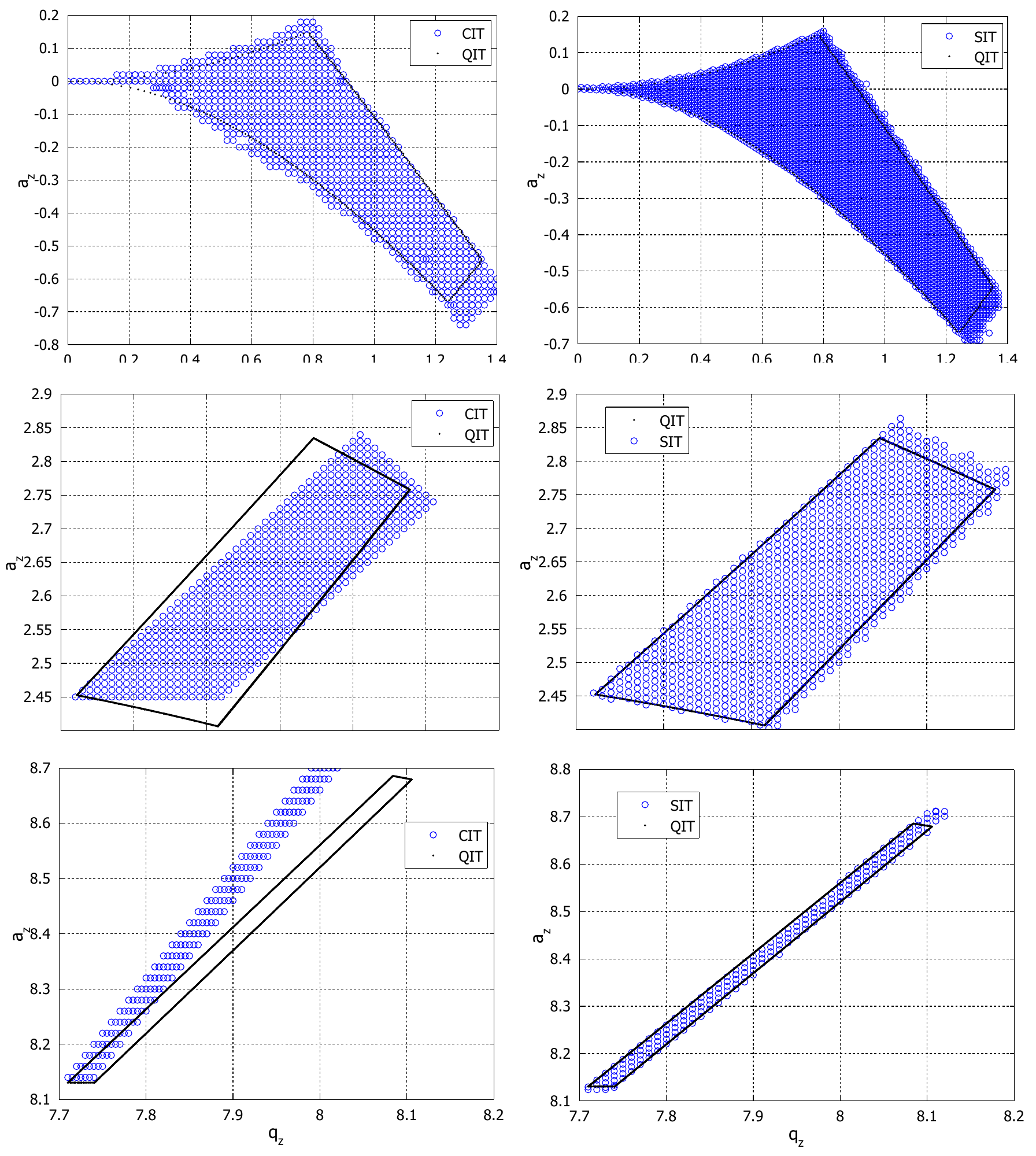}
\caption{Right: First, second and third stability regions (up to down respectively) calculated for SIT and compared with QIT. Left: First, second and third stability regions calculated for CIT \cite{noshad_numerical_2011} and compared with QIT \cite{noshad_computation_2009}.}
       \label{fig:5}

\end{figure}

\section{Computational Results}
\label{S:3}

We have solved the equations of motion using the fourth-order Runge-Kutta (RK4) method as well as the fifth-order Runge-Kutta (RK5) method with step-size control \cite{fehlberg_classical_1968}. Multipole components of the electric field up to $A_{12}$ are considered in the computations, and higher order terms have been set to zero. In our calculations, the radius of the trap was taken to be 15 mm. Initial velocity and position of the ions were considered to be 0 and  in both x and z coordinates, respectively. The trajectories of a charged particle inside the trap for typical values of Mathieu parameters $a$ and $q$ are depicted in Fig. \ref{fig:4}. It can be seen from the figure that for smaller values of $q$ the micromotions have a smaller amplitude relative to the main secular motion \cite{pearson_experimental_2006}.

Mathieu stability regions for an RF Paul trap are those values of $a$ and $q$ that lead to a stable solution of Mathieu equation. We computed three stability regions by solving the equations of motion numerically.Solutions that did not exceed the radius of the trap for up to 160 RF cycles were considered to be stable. The first, second and third stability regions are illustrated in Fig.\ref{fig:5}. The corresponding regions of an ideal QIT \cite{noshad_computation_2009} and a cylindrical ion trap (CIT) \cite{noshad_numerical_2011} are also included in the figures in order to be able to compare them at a glance. One can see that the difference between stability regions of QIT and CIT is larger than that of the QIT and our SIT for all three stability diagrams. The reason is the elimination of the electric octupole moment in our proposed spherical ion trap. The octupole moment has a noticeable effect on the fields, and is present for the CIT considered in \cite{noshad_numerical_2011}.

\section{Conclusions}
\label{S:4}
A highly symmetric radio-frequency quadrupole ion trap based on spherical geometry was presented. Optimization of the geometry along with the vanishing odd multipole components due to symmetry of the geometry made the potential field very similar to that of ideal hyperbolic Paul trap. The proposed spherical ion trap can be easily fabricated by considering merely $\alpha \simeq {{49}^{\circ }}$ as discussed in this article. Consequently, this trap is more suitable for miniaturization compared to the traditional hyperbolic Paul trap.

\section*{Acknowledgement}
This work was completed in part with resources provided by the High Performance Computing Research Center of Amirkabir University of Technology (\href{http://hpcrc.aut.ac.ir/}{hpcrc}).





\bibliographystyle{model1-num-names}
\bibliography{spherical-trap.bib}

\begin{thebibliography}{23}
\expandafter\ifx\csname natexlab\endcsname\relax\def\natexlab#1{#1}\fi
\providecommand{\bibinfo}[2]{#2}
\ifx\xfnm\relax \def\xfnm[#1]{\unskip,\space#1}\fi
\bibitem[{March(1997)}]{march_introduction_1997}
\bibinfo{author}{R.~E. March},
\newblock \bibinfo{title}{An introduction to quadrupole ion trap mass
  spectrometry},
\newblock \bibinfo{journal}{Journal of Mass Spectrometry} \bibinfo{volume}{32}
  (\bibinfo{year}{1997}) \bibinfo{pages}{351--369}.
\bibitem[{Cirac and Zoller(1995)}]{cirac_quantum_1995}
\bibinfo{author}{J.~I. Cirac}, \bibinfo{author}{P.~Zoller},
\newblock \bibinfo{title}{Quantum computations with cold trapped ions},
\newblock \bibinfo{journal}{Physical review letters} \bibinfo{volume}{74}
  (\bibinfo{year}{1995}) \bibinfo{pages}{4091}.
\bibitem[{Häffner et~al.(2008)Häffner, Roos, and
  Blatt}]{haffner_quantum_2008}
\bibinfo{author}{H.~Häffner}, \bibinfo{author}{C.~F. Roos},
  \bibinfo{author}{R.~Blatt},
\newblock \bibinfo{title}{Quantum computing with trapped ions},
\newblock \bibinfo{journal}{Physics Reports} \bibinfo{volume}{469}
  (\bibinfo{year}{2008}) \bibinfo{pages}{155--203}.
\bibitem[{Schmidt et~al.(2005)Schmidt, Rosenband, Langer, Itano, Bergquist, and
  Wineland}]{schmidt_spectroscopy_2005}
\bibinfo{author}{P.~O. Schmidt}, \bibinfo{author}{T.~Rosenband},
  \bibinfo{author}{C.~Langer}, \bibinfo{author}{W.~M. Itano},
  \bibinfo{author}{J.~C. Bergquist}, \bibinfo{author}{D.~J. Wineland},
\newblock \bibinfo{title}{Spectroscopy using quantum logic},
\newblock \bibinfo{journal}{Science} \bibinfo{volume}{309}
  (\bibinfo{year}{2005}) \bibinfo{pages}{749--752}.
\bibitem[{Rosenband et~al.(2008)Rosenband, Hume, Schmidt, Chou, Brusch, Lorini,
  Oskay, Drullinger, Fortier, Stalnaker, and
  {others}}]{rosenband_frequency_2008}
\bibinfo{author}{T.~Rosenband}, \bibinfo{author}{D.~B. Hume},
  \bibinfo{author}{P.~O. Schmidt}, \bibinfo{author}{C.~W. Chou},
  \bibinfo{author}{A.~Brusch}, \bibinfo{author}{L.~Lorini},
  \bibinfo{author}{W.~H. Oskay}, \bibinfo{author}{R.~E. Drullinger},
  \bibinfo{author}{T.~M. Fortier}, \bibinfo{author}{J.~E. Stalnaker},
  \bibinfo{author}{{others}},
\newblock \bibinfo{title}{Frequency ratio of al+ and hg+ single-ion optical
  clocks; metrology at the 17th decimal place},
\newblock \bibinfo{journal}{Science} \bibinfo{volume}{319}
  (\bibinfo{year}{2008}) \bibinfo{pages}{1808--1812}.
\bibitem[{Douglas et~al.(2005)Douglas, Frank, and Mao}]{douglas_linear_2005}
\bibinfo{author}{D.~J. Douglas}, \bibinfo{author}{A.~J. Frank},
  \bibinfo{author}{D.~Mao},
\newblock \bibinfo{title}{Linear ion traps in mass spectrometry},
\newblock \bibinfo{journal}{Mass Spectrometry Reviews} \bibinfo{volume}{24}
  (\bibinfo{year}{2005}) \bibinfo{pages}{1--29}.
\bibitem[{Louris et~al.(1987)Louris, Cooks, Syka, Kelley, Stafford~Jr, and
  Todd}]{louris_instrumentation_1987}
\bibinfo{author}{J.~N. Louris}, \bibinfo{author}{R.~G. Cooks},
  \bibinfo{author}{J.~E. Syka}, \bibinfo{author}{P.~E. Kelley},
  \bibinfo{author}{G.~C. Stafford~Jr}, \bibinfo{author}{J.~F. Todd},
\newblock \bibinfo{title}{Instrumentation, applications, and energy deposition
  in quadrupole ion-trap tandem mass spectrometry},
\newblock \bibinfo{journal}{Analytical Chemistry} \bibinfo{volume}{59}
  (\bibinfo{year}{1987}) \bibinfo{pages}{1677--1685}.
\bibitem[{Seidelin et~al.(2006)Seidelin, Chiaverini, Reichle, Bollinger,
  Leibfried, Britton, Wesenberg, Blakestad, Epstein, Hume, and
  {others}}]{seidelin_microfabricated_2006}
\bibinfo{author}{S.~Seidelin}, \bibinfo{author}{J.~Chiaverini},
  \bibinfo{author}{R.~Reichle}, \bibinfo{author}{J.~J. Bollinger},
  \bibinfo{author}{D.~Leibfried}, \bibinfo{author}{J.~Britton},
  \bibinfo{author}{J.~H. Wesenberg}, \bibinfo{author}{R.~B. Blakestad},
  \bibinfo{author}{R.~J. Epstein}, \bibinfo{author}{D.~B. Hume},
  \bibinfo{author}{{others}},
\newblock \bibinfo{title}{Microfabricated surface-electrode ion trap for
  scalable quantum information processing},
\newblock \bibinfo{journal}{Physical review letters} \bibinfo{volume}{96}
  (\bibinfo{year}{2006}) \bibinfo{pages}{253003}.
\bibitem[{Kielpinski et~al.(2002)Kielpinski, Monroe, and
  Wineland}]{kielpinski_architecture_2002}
\bibinfo{author}{D.~Kielpinski}, \bibinfo{author}{C.~Monroe},
  \bibinfo{author}{D.~J. Wineland},
\newblock \bibinfo{title}{Architecture for a large-scale ion-trap quantum
  computer},
\newblock \bibinfo{journal}{Nature} \bibinfo{volume}{417}
  (\bibinfo{year}{2002}) \bibinfo{pages}{709--711}.
\bibitem[{Wesenberg(2008)}]{wesenberg_electrostatics_2008}
\bibinfo{author}{J.~H. Wesenberg},
\newblock \bibinfo{title}{Electrostatics of surface-electrode ion traps},
\newblock \bibinfo{journal}{Physical Review A} \bibinfo{volume}{78}
  (\bibinfo{year}{2008}) \bibinfo{pages}{063410}.
\bibitem[{Ouyang et~al.(2004)Ouyang, Wu, Song, Li, Plass, and
  Cooks}]{ouyang_rectilinear_2004}
\bibinfo{author}{Z.~Ouyang}, \bibinfo{author}{G.~Wu},
  \bibinfo{author}{Y.~Song}, \bibinfo{author}{H.~Li}, \bibinfo{author}{W.~R.
  Plass}, \bibinfo{author}{R.~G. Cooks},
\newblock \bibinfo{title}{Rectilinear ion trap:  concepts, calculations, and
  analytical performance of a new mass analyzer},
\newblock \bibinfo{journal}{Analytical Chemistry} \bibinfo{volume}{76}
  (\bibinfo{year}{2004}) \bibinfo{pages}{4595--4605}.
\bibitem[{Gupta and Rao(2008)}]{gupta_numerical_2008}
\bibinfo{author}{A.~Gupta}, \bibinfo{author}{P.~M. Rao},
\newblock \bibinfo{title}{Numerical calculations of potential distribution in
  non-ideal quadrupole trap and simulations of anharmonic oscillations},
\newblock \bibinfo{journal}{Pramana} \bibinfo{volume}{70}
  (\bibinfo{year}{2008}) \bibinfo{pages}{457--470}.
\bibitem[{Sudakov(2001)}]{sudakov_effective_2001}
\bibinfo{author}{M.~Sudakov},
\newblock \bibinfo{title}{Effective potential and the ion axial beat motion
  near the boundary of the first stable region in a nonlinear ion trap},
\newblock \bibinfo{journal}{International Journal of Mass Spectrometry}
  \bibinfo{volume}{206} (\bibinfo{year}{2001}) \bibinfo{pages}{27--43}.
\bibitem[{Kim et~al.(2010)Kim, Herskind, Kim, Kim, and
  Chuang}]{kim_surface-electrode_2010}
\bibinfo{author}{T.~H. Kim}, \bibinfo{author}{P.~F. Herskind},
  \bibinfo{author}{T.~Kim}, \bibinfo{author}{J.~Kim}, \bibinfo{author}{I.~L.
  Chuang},
\newblock \bibinfo{title}{Surface-electrode point paul trap},
\newblock \bibinfo{journal}{Physical Review A} \bibinfo{volume}{82}
  (\bibinfo{year}{2010}) \bibinfo{pages}{043412}.
\bibitem[{Major et~al.(2006)Major, Gheorghe, and Werth}]{major_charged_2006}
\bibinfo{author}{F.~G. Major}, \bibinfo{author}{V.~N. Gheorghe},
  \bibinfo{author}{G.~Werth}, \bibinfo{title}{Charged particle traps: physics
  and techniques of charged particle field confinement},
  volume~\bibinfo{volume}{37}, \bibinfo{publisher}{Springer},
  \bibinfo{year}{2006}.
\bibitem[{Griffiths and College(1999)}]{griffiths_introduction_1999}
\bibinfo{author}{D.~J. Griffiths}, \bibinfo{author}{R.~College},
  \bibinfo{title}{Introduction to electrodynamics},
  \bibinfo{publisher}{Prentice hall Upper Saddle River, {NJ}},
  \bibinfo{year}{1999}.
\bibitem[{Jackson(1999)}]{jackson_classical_1962}
\bibinfo{author}{J.~D. Jackson}, \bibinfo{title}{Classical electrodynamics},
  \bibinfo{publisher}{Wiley New York etc.}, \bibinfo{year}{1999}.
\bibitem[{Wang et~al.(1993)Wang, Franzen, and Wanczek}]{wang_non-linear_1993}
\bibinfo{author}{Y.~Wang}, \bibinfo{author}{J.~Franzen}, \bibinfo{author}{K.~P.
  Wanczek},
\newblock \bibinfo{title}{The non-linear resonance ion trap. part 2. a general
  theoretical analysis},
\newblock \bibinfo{journal}{International journal of mass spectrometry and ion
  processes} \bibinfo{volume}{124} (\bibinfo{year}{1993})
  \bibinfo{pages}{125--144}.
\bibitem[{Plass et~al.(2003)Plass, Li, and Cooks}]{plass_theory_2003}
\bibinfo{author}{W.~R. Plass}, \bibinfo{author}{H.~Li}, \bibinfo{author}{R.~G.
  Cooks},
\newblock \bibinfo{title}{Theory, simulation and measurement of chemical mass
  shifts in {RF} quadrupole ion traps},
\newblock \bibinfo{journal}{International Journal of Mass Spectrometry}
  \bibinfo{volume}{228} (\bibinfo{year}{2003}) \bibinfo{pages}{237--267}.
\bibitem[{Noshad and Kariman(2011)}]{noshad_numerical_2011}
\bibinfo{author}{H.~Noshad}, \bibinfo{author}{B.-S. Kariman},
\newblock \bibinfo{title}{Numerical investigation of stability regions in a
  cylindrical ion trap},
\newblock \bibinfo{journal}{International Journal of Mass Spectrometry}
  \bibinfo{volume}{308} (\bibinfo{year}{2011}) \bibinfo{pages}{109--113}.
\bibitem[{Noshad and Doroudi(2009)}]{noshad_computation_2009}
\bibinfo{author}{H.~Noshad}, \bibinfo{author}{A.~Doroudi},
\newblock \bibinfo{title}{Computation of five stability regions in a quadrupole
  ion trap using the fifth-order runge–kutta method},
\newblock \bibinfo{journal}{International Journal of Mass Spectrometry}
  \bibinfo{volume}{281} (\bibinfo{year}{2009}) \bibinfo{pages}{79--81}.
\bibitem[{Fehlberg(1968)}]{fehlberg_classical_1968}
\bibinfo{author}{E.~Fehlberg}, \bibinfo{title}{Classical fifth-, sixth-,
  seventh-, and eighth-order Runge-Kutta formulas with stepsize control},
  \bibinfo{publisher}{National Aeronautics and Space Administration; for sale
  by the Clearinghouse for Federal Scientific and Technical Information,
  Springfield, Va.}, \bibinfo{year}{1968}.
\bibitem[{Pearson et~al.(2006)Pearson, Leibrandt, Bakr, Mallard, Brown, and
  Chuang}]{pearson_experimental_2006}
\bibinfo{author}{C.~E. Pearson}, \bibinfo{author}{D.~R. Leibrandt},
  \bibinfo{author}{W.~S. Bakr}, \bibinfo{author}{W.~J. Mallard},
  \bibinfo{author}{K.~R. Brown}, \bibinfo{author}{I.~L. Chuang},
\newblock \bibinfo{title}{Experimental investigation of planar ion traps},
\newblock \bibinfo{journal}{Physical Review A} \bibinfo{volume}{73}
  (\bibinfo{year}{2006}) \bibinfo{pages}{032307}.

\end{thebibliography}







\end{document}